\documentclass[twocolumn,showpacs,preprintnumbers,amsmath,amssymb]{revtex4}
\usepackage{graphicx, amsmath, amssymb}
\begin{document}

\title{Fluxon-based generation of graph states in Josephson qubits}

\author{Toshiyuki Fujii$^1$}

\author{Masaru Fukunaga$^2$}
\author{Noriyuki Hatakenaka$^1$}

\affiliation{$^1$Graduate School of Integrated Arts and Sciences, 
  Hiroshima University, Higashi-Hiroshima, 
  739-8521, Japan}
  
\affiliation{$^2$Department of Physics,
  Hiroshima University, Higashi-Hiroshima, 
  739-8526, Japan}

\date{\today}


\begin{abstract}
Graph states are a special kind of multiparticle entangled state with great potential for applications 
in quantum information technologies, especially in measurement-based quantum computers. 
These states cause significant reductions of the number of qubits needed for a given computation, 
leading to shorter execution time. 
Here we propose a simple scheme for generating such graph states by using special gate operations, 
i.e., control-phase and swap gate operations, inherent in superconducting quantum nanocircuits. 
\end{abstract}
\maketitle

Entanglement is at the heart of quantum mechanics and is used in quantum information technologies. 
Graph states \cite{Schlingemann, Clever, 1way, experiment1,   pra69-062311} are special kinds of multiparticle entangled states 
that correspond to mathematical graphs where the vertexes take the role of qubits 
and the edges represent Ising-type interactions between pairs of qubits, 
and serve as the central resource in quantum information technologies 
such as quantum error correction \cite{Schlingemann}, multi-party quantum communication \cite{Clever}, 
and most prominently in measurement-based quantum computation \cite{1way,experiment1}.
Graph states can significantly reduce the number of qubits needed for a given computation,
resulting in the anomalous reduction of computation processes \cite{pra69-062311}. 
This is due to extra flexibility in the entanglement rich structure of graph states 
that results from the absence of limitation as regards connecting partners.
In other words, graph states require far fewer qubits to implement the same measurement-based  
quantum computations compared with cluster states \cite{QCc1, QCc2, QCc3, 1stepGen, quant-ph0703090}, which are regarded as a sub-class of graph states. 
This allows us to design the circuits in more compact forms. 

However, many physical systems that can generate graph states are limited to cluster state generation, 
because the physical qubit interactions are limited to some kind of nearest-neighbor form. 
In this Letter, we propose a simple scheme for generating $N$-connected graph states by using fluxon in Josephson transmission line (JTL). 

The graph state for $N$ qubits $|G\rangle$ is defined as 
\begin{equation}
|G\rangle = \prod_{(i,j)} C(z)^{(i,j)} |+\rangle^{\otimes N}, 
\end{equation}
where $(i,j)$ stands for two connected vertexes in the graph, and $|{+}\rangle=( |{0}\rangle+ |{1}\rangle)/\sqrt{2}$. 
$C(z)^{(i,j)} $ denotes a controlled phase (CP) operator that reverses the phase 
when two qubits  $i$ and $j$ are in the  $|11\rangle$ state.
Graph states are then entangled quantum states that exhibit complex structures of genuine multi-particle entanglement. 
Of these, an {\it $N$-connected graph state } is a state that corresponds to a graph 
with an $(i, j)$-path for any two vertexes $(i, j)$ among $N$ vertexes. 
Any graph can be produced from the $N$-connected graph by cutting unwanted edges. 
Here we propose a simple scheme for generating an $N$-connected graph in superconducting quantum nanocircuits. 

\begin{figure}[here]
 \begin{center}
 \includegraphics[width=7cm]{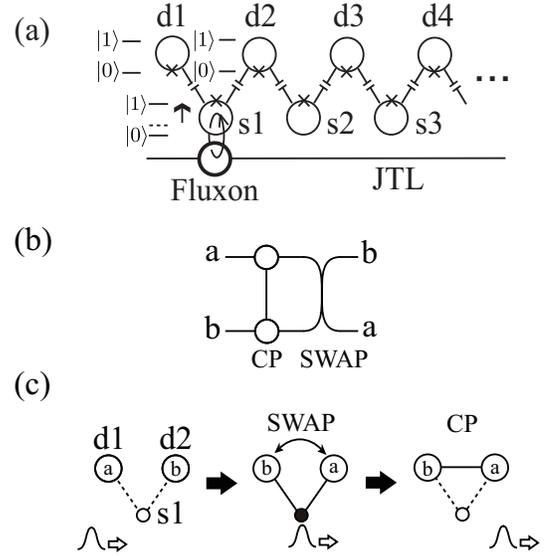}
 \end{center}
 \caption{ 
(a) Flux qubit chain.  The energy-level separation of s1 is shifted by fluxon. 
(b) Quantum circuit diagram of the CP+SWAP operation.
(c) The operation $\tilde U$ expressed by Eq. \eqref{is2} results in 
the connection of data qubit $a$ and $b$ with swapping by a {\it one-way } fluxon propagation. }
 \label{fig:1}
\end{figure}

The system we are considering is shown in Fig. \ref{fig:1} (a). 
It is composed of a Josephson transmission line  
and a zigzag chain of bipartite superconducting flux qubits 
with an alternating arrangement, i.e., one qubit functions as a data qubit that are labeled d$i$, 
while its nearest neighbor labeled s$i$ functions as a switch between data bits \cite{matsuo}.
The energy-level separations of all the data qubits are assumed to be equal 
and very different from those of switch qubits. 
Thus, the data qubits are initially decoupled from each other 
in this system because of the off-resonance between nearest neighbor switch qubits.

The coupling of two data qubits d$1$ and d$2$ is realized via a third (switch) qubit s$1$, which is controlled 
by a fluxon motion, i.e., 
the qubit-qubit interaction is activated when the fluxon induces an energy-level shift 
equal to the energy-level separation of the data qubit so as to resonate energetically among three qubits. 
During the resonance, the state vector of three qubits $|\Psi (t) \rangle$ evolves through the relation 
$|\Psi (t) \rangle = U(t) |\Psi(0)\rangle $, with $U(t)$ being a time-translational operator for three qubits s$1$, d$1$ and d$2$. 
A significant example for quantum computation is the transfer of information in the qubit chain \cite{matsuo}. 
In particular, the state vector starting from the switch qubit state $|{0}\rangle_{s1}$ is decoupled 
into a data-qubit state $\tilde U |{\psi (0)}\rangle$ and a switch-qubit state $|{0}\rangle_{s1}$ with $\pi$ pulse 
application, i.e., $ U(t_\pi) \{ |{\psi (0)}\rangle |{0}\rangle_{s1} \} = \{ \tilde U |{\psi (0) }\rangle \} |{0}\rangle_{s1}$ 
where $t_\pi = \hbar \pi /g \sqrt{2}$ with $g$ being a coupling constant. 
The effective time-translational operator for the data-qubit system $ \tilde U$ 
is then expressed in a matrix form as
\begin{equation}
 \tilde U =
- \begin{pmatrix}
-1&& 0 &&0&& 0 \\
0&&1 && 0 &&0 \\
0 && 0 &&1 &&0 \\
0&&0 &&0&& 1 \\
\end{pmatrix} \begin{pmatrix}
1&& 0 &&0&& 0 \\
0&&0 && 1 &&0 \\
0 && 1 &&0 &&0 \\
0&&0 &&0&& 1 \\ 
\end{pmatrix}, \label{is2}
\end{equation}  
where the basis is ordered as $ |00\rangle$, $ |01\rangle$, $ |10\rangle$, $ |11\rangle$ 
with $|nm\rangle\equiv |n\rangle_{d1 } \otimes |m\rangle_{d2} $. 
This is regarded as a gate composed of a joint-phase (JP) gate, 
which reverses its sign only in the $|{00}\rangle$  state  and a SWAP gate. 
This is sometimes called a JP+SWAP gate \cite{Yung}. 
Note that the JP gate is equivalent to the CP gate, which reverses the sign of the state $|11\rangle$, 
if the bases of the data qubits are renamed as $|0 \rangle \rightleftharpoons |1 \rangle$. 
Hereafter, we exchange the two bases and regard the JP gate as the CP gate. 
The circuit diagram of the operation is shown in Fig. \ref{fig:1} (b). 
Figure \ref{fig:1} (c) shows blocked processes of a CP+SWAP operation. 
The switch qubit is changed from off-resonant (white circle) to on-resonant (black circle) 
when a fluxon approaches. 
The vertexes denoted by $a$ and $b$ are connected by the edge represented by a solid line. 
Note that the location of vertexes are exchanged each other. 
This mobile nature functions effectively for connecting all the vertexes. 

\begin{figure}
 \begin{center}
 \includegraphics[width=7cm]{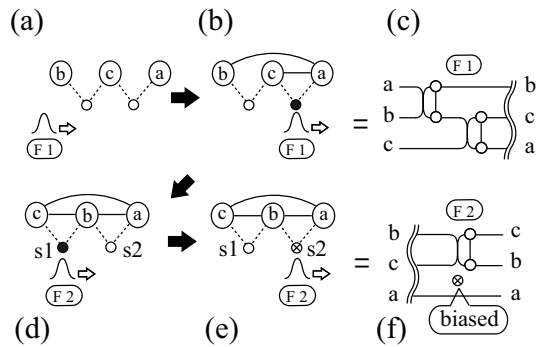}
 \end{center}
 \caption{
(a) Initial preparation; empty graph state. 
(b) The graph state caused by the passage of the first fluxon (F1). 
(c) The circuit diagram of operation by a  propagation of the F$1$ for 3 qubits. 
(d) The second fluxon  first biases to the switch qubit 1  and operates the SWAP  to vertexes $b $ and $c$. 
(e) The bias to  switch qubit $2$ has no effect because of the extra bias. 
The generated state is a $3$-connected graph state.
(f) Circuit diagram of operation by F2. }
 \label{fig:2}
\end{figure}

Now let us consider the $N$-connected graph state. For simplicity, we first consider the $N=3$ case. 
Figure \ref{fig:2} (a) shows that three data qubits are initially prepared 
in the state $|{+}\rangle=(|{0}\rangle+|{1}\rangle)/\sqrt{2}$, i.e., $|+\rangle^{\otimes 3} $. 
The first fluxon creates a graph as shown in Fig. \ref{fig:2} (b). 
The vertex $a$ moves from data qubit 1 to data qubit 3, 
together with the exchange of vertexes $b$ and $c$. 
In other words, cyclic permutation occurs at three vertexes. 
At the same time, CP operations are performed at any pair of vertexes 
between vertex $a$ and the other vertexes, i.e., $b$ and $c$.
The corresponding circuit diagram of this operation is shown in Fig. \ref{fig:2}(c).

\begin{figure}
 \begin{center}
 \includegraphics[width=6.5cm]{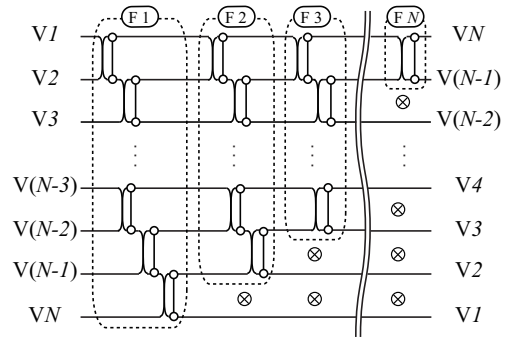}
 \end{center}
 \caption{Circuit diagram of generation of $N$-connected graph state.
 \label{fig:3} }
\end{figure}

To form the three-connected graph state, an edge should be introduced between  the $b$ and $c$ vertexes. 
The second fluxon can establish the $b-c$ edge without cutting other connected edges 
if the extra flux bias is applied to switch qubit $2$ before the second fluxon enters the system. 
The extra bias works to produce the off-resonance between vertexes $a$ and $b$ even if the fluxon reaches it. 
Under this condition, the resonant oscillation only occurs between vertexes $b$ and $c$  
when the second fluxon reaches the switch qubit $1$. 
Vertexes $b$ and $c$ exchange positions and form an edge as shown in Fig. \ref{fig:2} (d).
Figure \ref{fig:2} (e) shows the generated three-connected graph state. 
All the vertexes are connected to each other. 

The general $N$-connected graph state can be generated by extending this method as shown in Fig. \ref{fig:3}. 
In this scheme, $N-1$ fluxon propagations are required.
The first fluxon (F$1$) connects vertex V$1$ and the rest of the vertexes, i.e., V$2, \cdots,$ and V$N$. 
Before the passage of the second fluxon (F$2$), 
we apply the extra bias to the $(N-1)$-th (the rightmost) switch qubit 
to avoid cutting the edge between V$2$ and V$1$ established by the first fluxon. 
Then the second fluxon connects V$2$ and the rest of vertexes except for V$1$. 
In the same way, the third fluxon connects V$3$ and rest of the vertexes V$4, \cdots,$ and V$N$ 
when the switch qubit $(N-1)$ and $(N-2)$ are applied with an extra bias flux for the reason described above. 
Similar procedures are repeated until all $(N-1)$-time passages of a fluxon. 
As a result, the $N$-connected graph state is generated on the data qubits.
 
\begin{figure}
\begin{center}
\includegraphics[width=7cm, clip]{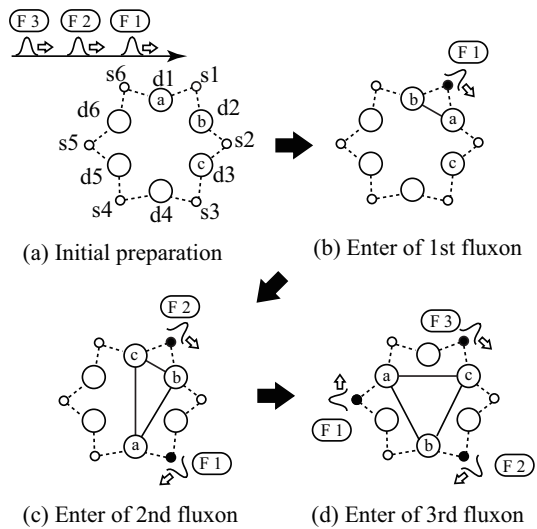}
\end{center}
\caption{Second scheme without extra magnetic bias switching.}
 \label{fig:4}
\end{figure}
The scheme proposed above requires extra bias switching processes, leading to a longer execution time. 
This is not good for a system with a shorter decoherence time, especially in solid-state qubits. 
Now let us consider $N$-connected graph state generation without such extra operations 
by using a circular flux qubit chain. 
For simplicity, we again consider an $N=3$ case. 
It is straightforward to extend the scheme to more general $N$-connected graph state generation. 
In this scheme, we prepare a circular flux qubit chain with $2N=6$ data qubits as shown in Fig. \ref{fig:4} (a).
The data qubis d$1$, d$2$ and d$3$ containing vertexes $a$, $b$, and $c$ are prepared in the state $|+\rangle$. 
The other three data qubits d$4$, d$5$ and d$6$ are prepared in the state $|0\rangle$ 
and work as a ``spacer" between the vertexes. 
A circular Josephson transmission line runs along the this flux qubit chain. 

Three fluxons are successively introduced into the circular JTL.
Figure \ref{fig:4} (b) shows that the first fluxon comes into the circular JTL 
and biases to the switch qubit 1 (s1).
This connects vertexes $a$ and $b$ and simultaneously exchanges these two vertexes. 
The fluxon sequentially biases to the next switch qubit (s$2$) and connects $a$ and $c$.
The second fluxon (F$2$) enters the circular JTL 
when F1 reaches  s$3$, 
(see Fig. \ref{fig:4} (c)). 
F$2$ forms an edge between $b$ and $c$, 
while the first fluxon F$1$ exchanges vertex $a$ and spacer located at d$3$ and d$4$, respectively.
At this time, the CP gate operation is also performed 
between the spacer and vertex $a$ along with their exchange. 
However, the CP gate that only flips the phase of the state $|11\rangle$ does not change the state of  
  the spacer and vertex $a$ 
since the spacer is prepared in the state $|0\rangle$. 
Thus, F$1$ only causes the vertex $a$ to slide from d$3$ to d$4$.
The resulting state is a three-connected graph state as shown in Fig. \ref {fig:4} (c). 
Figure \ref{fig:4} (d) shows a state when the third fluxon (F3) comes into the circular JTL. 
Each fluxon in the circular JTL simultaneously exchanges the vertex and neighboring spacer.
This operation only slides the vertex into its neighboring data qubit location. 
Thus the edges are not cut by this operation.  
This scheme can be extended to general $N$-connected graph state generation 
by using $N$ fluxons and $2N$ data qubits.

In summary, we proposed two schemes for generating the $N$-graph state in a flux qubit chain.
The first scheme generates the $N$-connected graph state in a line-arranged flux qubit chain 
by means of an $N-1$ fluxon and external bias switching processes. 
In contrast, the second scheme 
generates the $N$-connected graph state in the circularly-arranged flux qubit chain 
by using an $N$ fluxon without external bias switching processes.
The proposed fluxon-based gate control simplifies the switching of inter-qubit coupling and gate operations. 
This scheme is applicable to the generation of graph states with an arbitrary configuration, 
and is effective in quantum error correction and quantum key distribution. 

This work was supported in part by KAKENHI (Nos. 18540352 and 195836) from MEXT of Japan.

\end{document}